**Michal Natorski**

Maastricht University

UNU-MERIT, United Nations University

**E-mail:** michal.natorski@maastrichtuniversity.nl

**ORCID:** 0000-0002-2736-1709

# Compromise in Multilateral Negotiations and the Global Regulation of Artificial Intelligence

**Abstract**

As artificial intelligence (AI) technologies spread worldwide, international discussions have increasingly focused on their consequences for democracy, human rights, fundamental freedoms, security, and economic and social development. In this context, UNESCO's Recommendation on the Ethics of Artificial Intelligence, adopted in November 2021, has emerged as the first global normative framework for AI development and deployment. The intense negotiations of every detail of the document brought forth numerous controversies among UNESCO member states. Drawing on a unique set of primary sources, including written positions and recorded deliberations, this paper explains the achievement of global compromise on AI regulation despite the multiplicity of UNESCO member-state positions representing a variety of liberal and sovereignist preferences. Building upon Boltanski's pragmatic sociology, it conceptualises the practice of multilateral negotiations and attributes the multilateral compromise to two embedded therein mechanisms: Structural normative hybridity and situated normative ambiguity allowed to accomplish a compromise by linking macro-normative structures with situated debates of multilateral negotiations.

# 1. Introduction

Artificial Intelligence (AI) systems increasingly assist and even replace human decisions in military, judicial, policing, healthcare, transport, financial, education, or social protection fields[1]. The regulation of AI technologies has recently become a worldwide priority given the unintended and unpredictable consequences of uncontrolled diffusion of AI related to democracy, human rights, fundamental freedoms, security, and economic and social development[2]. The most advanced attempts at international regulation of AI take place in the framework of regional international organisations such as the Organization for Economic Cooperation and Development, the Council of Europe, G7 and the European Union[3]. In this context, the Recommendation on the Ethics of Artificial Intelligence adopted by all 193 UNESCO member states as "an international standard-setting instrument" became the first instance of global regulation of AI[4].

To reach such a global agreement, since March 2020, UNESCO member states and stakeholders debated the outline of the Recommendations. After multi-stakeholder consultations, the UNESCO Secretariat presented, in March 2021, a draft of the Recommendations submitted to the final phase of intergovernmental negotiations spanning from April to June 2021[5]. During this stage of negotiations, liberal-oriented states prioritised the need to protect universal human rights, gender equality and a multi-stakeholder approach in the implementation of AI policies. However, many sovereignist-oriented states simultaneously emphasised the protection of state sovereignty and cultural specificities in implementing AI policies. Despite tense negotiations confronting different preferences, all UNESCO states adopted the Recommendations during the UNESCO General Conference in November 2021[6]. The agreement links liberal priorities of universal values of the protection of human rights based on international law and the promotion of diversity and inclusiveness with the sovereigntist position of state control in the implementation of this agenda. Therefore, the adoption of UNESCO Recommendations on the Ethics of AI seems to challenge the image of an apparent crisis of the international liberal order and fragmentation of the multilateral system[7], which is generally attributed to the cleavages between liberalism and Westphalian sovereignty principles[8], liberal democracies and authoritarian populism[9] or communitarianism and cosmopolitanism philosophies[10].

Such clashes between liberal and sovereigntist positions have already defined the multilateral debate about data governance[11]. Data constitutes the backbone of AI systems; therefore, the accomplished agreement on the UNESCO AI Recommendations across this cleavage is empirically puzzling. It shows that cooperation on crucial global issues can develop despite the confrontation between the liberal supporters of multilateral institutions and its challengers protecting state sovereignty. This is also theoretically puzzling for the mainstream analytical approaches to multilateral negotiations frequently portrayed along the bargaining-arguing dichotomy. From a rationalist perspective, the enthusiast negotiation of non-binding ethical principles is at odds with the assumption of the irrelevance of moral discourse among cost-benefit calculating interest-maximizing parties. All states have adopted the UNESCO AI Recommendations appealing to international norms and law in their positions to justify either liberal or sovereignist stances. As a result, even following rationalist reasoning, to produce an agreement, parties accommodated competing standards of justice to satisfy the demand for equity on all sides[12].

From a normative perspective, despite the document's "legally non-binding" character, the intense negotiations might be attributed to the states' awareness that this "soft law" AI regulation might lead to "hard" normative commitment establishing a global precedent for future normative and legal actions[13] requiring states' compliance with them[14]. However, the dynamics of these negotiations contradict the assumption of the arguing approach, which explains the outcomes of multilateral negotiations by the shifts in states' positions due to persuasion and normative rationality leading to normative convergence[15]. During the examined negotiations, no state unequivocally shifted its initial positions or joined a group of countries justifying different approaches. On the contrary, all states followed their general positions during the negotiation until the last hour of deliberation, and despite the persistent diversity of positions, they achieved a compromise. Consequently, the arguing approach to negotiations does not fully account for arriving at a compromise despite the persistent multiplicity of normative views. It overlooks the process of the emergence of international norms' complexity and ambiguity[16] and how they are agreed upon among contradictory justificatory claims during multilateral negotiations[17].

To a different degree, both mainstream approaches recognise the relevance of normativity during multilateral deliberations but do not offer a persuasive answer to the following question: How is a compromise achieved in multilateral negotiations amid normative multiplicity represented by states? To address this question, this paper adopts the practice theory approach as an alternative to rationalist and norm-oriented theories that assume a specific normative rationality of actors[18]. In more detail, it follows the approach developed by Luc Boltanski to disclose the normativity elements in the everyday performance of agents and capture the accomplishment of normative compromises in multilateral negotiations. First, contrary to mainstream theories, this approach allows conceptualising and empirically illustrating the practices of multilateral negotiating to reach a compromise within the landscape of normative multiplicity rather than a principled confrontation between clearly delimited normative cleavages. Second, this approach suggests two mechanisms for reaching a compromise among a diversity of justificatory claims. The first mechanism is crafting the structural normative hybridity reflecting the amalgamation of different normative positions within the negotiated text. The second mechanism of situated normative ambiguity is the meticulous re-drafting of the contentious words, concepts, and terms to convey ambiguous meanings allowing for a normative interpretation acceptable to different actors.

The following section conceptualises the practices of multilateral negotiations and the two embedded mechanisms for achieving a compromise. The third section outlines the multiplicity of normative priorities in the UNESCO member states positions towards the negotiated text to identify shared elements constituting a background for a compromise. The fourth and fifth sections analyse the two mechanisms of achieving a compromise while negotiating the draft of the text. The final section summarises the findings and outlines future avenues for research on the practice of multilateral compromises.

## 2. The practice of multilateral negotiation and the mechanisms of compromises

The analytical framework outlines the distinctive premises of Boltanski's framework to conceptualise the practice of multilateral negotiations and the two mechanisms to achieve a normative compromise. It argues that the mechanisms of structural normative hybridity and situated normative ambiguity are embedded in the process of addressing normative controversies. Finally, this section outlines the methodology employed to analyse the collected data.

### 2.1. Orders of worth and normative multiplicity

Practices can be defined as socially meaningful and organised patterns of activities embedded within normative contexts[19]. Following the shared elements of the international practice theories approach[20], the practice of multilateral negotiating might be defined as a rule-based *process* of (re)presenting different views to reach an agreement *materialised* in a text. Negotiating is *performed* by agents equipped with their *practical knowledge* of translating abstract *multiple* normative positions into *empirical* terminologies. The outcome of this practice is a *collective* compromise satisfying different normative positions.

To conceptualise achieving a compromise in the practice of multilateral negotiations as defined above, this paper employs the pragmatic sociology of Luc Boltanski and his collaborators[21]. From this perspective, negotiating might be perceived as a continuous stream of routinised actions embedded in situations of normative controversies[22]. Situations are open to interpretation and require formulation, modification and reformulation of goals and means[23]. Hence, negotiating includes *performances* of actions, such as delegations' preparing and (re)presenting the positions on the negotiated matter, participating in the meetings, following the other participants' positions, proposing solutions to controversies, and agreeing on outcomes.

The concept of order of worth underpinning Boltanski's approach offers an avenue to convey the normative *multiplicity* represented by different actors during multilateral negotiations. Orders of worth are co-existing normative moral principles concerning the ordinary sense of justice and are employed by actors to assess the value of practices. They are plural systems not reducible to each other and can legitimately co-exist in one social space[24]. They offer different normative references invoked by actors to judge specific actions in different social spheres "to make sense of their everyday situations and reach situated agreements under a 'higher common principle'"[25].

All orders of worth follow the same model with "a common underlying structure or, if you like, grammar"[26] reflected in a system of conceptual tools and associated vocabulary. They are constituted around the assumption of *the common good in a political community,* offering rationales for determining the unequal moral status of the community member[27]. In each order of worth, the *principle of equivalency* provides a general standard to evaluate all actions, things, and persons in specific situations. For example, Boltanski distinguished the principle of competition to define the order of worth of the market and the principle of efficiency in the industrial order of worth based on technological and scientific advancements. Therefore, by claiming the value of people in their situated activities, each principle of equivalency orders hierarchical relations of superiority and inferiority. This principle can be used to characterise virtues and deficiency of subjects and objects and offers references to interpret and classify controversial situations and test moral status[28].

The orders of worth can be used "as a general heuristic for reconstructing repertoires of evaluation as shared moral narratives" in different domains of global society[29]. Coexisting normative orders might be mobilised in multilateral negotiations when the negotiated matters are interpreted from different normative perspectives represented by states. But this approach does not imply a set of abstracted normative positions but follows the *empirical* analysis of actors' "interpretation of the world"[30]. The *empirical* conceptualisations of the normative positions combine different sources. For example, Boltanski distinguished seven orders of worth by validating their core elements derived from philosophical debates[31] and French management literature[32] with the empirical analysis of action-oriented guidebooks intended to instruct everyday human actions.

Following the premises of plurality and contextual character of the orders of worth[33], this paper defines the normative positions towards multilateral cooperation in two empirical steps. The first step adopts the liberal and sovereignist positions as the dominant cleavage in international affairs regarding multilateral cooperation. These ideas originate in Kantian and Hobbesian philosophies, inspiring the thinking about the common good in international affairs around the *principles of universalism* and *particularism*, respectively. They share several common normative elements, such as the self-determination of states and non-inferences in internal affairs. At the same time, they differ as regards the role of human rights and the responsibility for collective intervention to protect liberal values[34] as well as the scope of the state control regarding the authority of international institutions and private actors[35]. These general approaches were recently reinterpreted as communitarian and cosmopolitan normative positions towards the state's role in globalisation. The cosmopolitan stance defends open state borders to stimulate free international interactions, powerful international institutions, the individuals as the point of reference and the justification of the universal rights of all human beings to uphold their freedoms. The communitarian perspective supports clear demarcation of borders, the defence of state sovereignty against international authorities, and the prioritisation of cultural and national communities' freedoms to realise self-determination[36]. In the second step of the analysis, the general categories of the orders of worth gain specific expressions from the empirical materials, in this case, the general state positions presented during the opening of negotiations. In this way, it does not assume an ideal type of liberal and sovereignist position on regulating AI but generalises it from the empirical materials following the analytical categories defined in Table 1. It allows us to find differences and shared elements between these orders of worth.

**Table 1. The structure of the orders of worth**

| Organizing dimensions | Liberal | Sovereigntist |
|---|---|---|
| **Principle of equivalency** | Universalism | Particularism |
| **Common good** | Individual freedoms from oppression | Communitarian freedoms to self-determination |
| **Community** | Humanity formed by individuals | Collectives formed around states and cultures |
| **Virtues** | Universal ethics to protect common values | Different cultural values |
| | Equal and inclusive for all people and issues (gender equality) | Respectful and inclusive to cultural and state diversity |

|  | Global institutions' authority and multi-stakeholder processes | Sovereignty and state authority |
|---|---|---|
|  | Global scope and agenda | Limited scope and agenda |
|  | Common standards for implementation | Non-binding and flexible for implementation |
| **Deficiencies** | Discrimination, exclusion, and inequality of individuals | Discrimination, exclusion and inequality of states and communities |
|  | Issue exclusion from regulation | Broad issue inclusion |
|  | Threats and risks for individuals | Threats and risks for communities |
| **Repertoire of valued objects** | Instrument for universal protection: universal human rights law, international law, global institutions, innovation for human development | Instruments for community protection: State law, national strategies, cultural values, innovation for community development |

Source: Own elaboration

### 2.2. Tests and normative compromises

A landscape of normative multiplicity is *performed* by statements and comments during negotiations. They focus on commenting and amending a draft text by simultaneously interpreting it from the perspective of general normative references and converting normative preferences into a specific text. This process transcends the dichotomy of everyday activities and abstract norms since "in practices, micro-(situations) and macro-perspectives (orders of worth) play together when actors attempt to cope with conflicts in everyday life"[37].

Tests are the *performances* when the orders of worth are exposed and justified in real-life situations such as multilateral negotiations. It reflects the examination of the specific understandings of the text concerning different normative references. Involved actors refer to their conceptions of the common good in society and associated principles of equivalency. These critical moments can be portrayed as "*the scene of a trial*, in the course of which actors in a situation of *uncertainty* proceed to *investigations*, record their *interpretations* of what happens in *reports*, establish *qualifications* and submit to *tests*"[38]. During a test, actors try to achieve an agreement on the qualification of the situation by confronting different criteria of equivalence of orders of worth to evaluate their worth in each case. It leads to a continuous process of "normative ordering"[39].

Boltanski distinguished the *tests of truth, reality and existence* that might be used to legitimise, reinterpret, or change the existing order of worth respectively[40]. This categorisation resembles three situations during multilateral negotiations and the possible role of text amendments. In *the truth test*, the consonance between the wordings and the normative references will be recognised during the debate, and proposed amendments will eventually reaffirm it. In the *reality test*, specific wording can be criticised, given the perceived mismatch with normative orders. The *reality test* leads to a collective search for a compromise in which an amendment of language can be proposed to improve the consonance with orders of worth. Finally, *the existential test* contests the normative approach reflected in specific wordings. Presented amendments claim normative specificity and limit flexible understandings. As a result, they will likely be refused as they obstruct consensus-building to reach an agreement with other normative orders.

The presentation and negotiations of amendments rely on the *practical knowledge* of the involved actors. They are equipped with critical capacities and reflexive competence to face controversies and reach contingent agreements based on an ordinary sense of justice and the contribution to the common good[41]. Negotiation implies recognising "an inclusive, institutionalised and principled form of political dialogue"[42]. Therefore, when discussing a text from their normative perspective, negotiators possess the skills to convey arguments to reach a compromise to be accepted by different normative positions. Practical knowledge involves linguistic skills creatively conveying normative concerns in a particular situation[43].

This practical knowledge is learnt *collectively* during the debates, and reaching a compromise is the collective process of meaning-making in a specific context[44]. During the negotiation, agents learn what kind of utterances are expected and acceptable to satisfy different normative positions to reach a compromise. They also understand what will be inadmissible. Boltanski suggests that a compromise might be achieved without undermining the validity of disputed orders of worth, assuming that "people can reach justifiable agreements despite the availability of multiple principles of agreement and can do so without acknowledging a relativism of values"[45]. A compromise aims at suspending the clash between different orders of worth, by seeking the general interest universally applicable and by defining a common good which includes various forms of worth. It can also be based on "a principle [of equivalence] that can take judgments based on objects stemming from different worlds and make them compatible"[46]. Hence, it can involve trade-offs between the different orders since pragmatic actors usually draw from elements of different orders[47].

During the negotiations, the text of a document becomes a key point of reference, whereas national interests are defined in textual edits and the compromise is achieved through "aesthetically circumscribed drafting exercises"[48]. Following Boltanski, a compromise might be *materialised* in a text in two ways.

First, the structural normative hybridity consists of a skilful merging of the elements of orders of worth into a composite arrangement, whereas "beings that matter in different worlds [orders] are maintained in presence, but their identification does not provoke a dispute"[49]. A compromise in such a situation is underpinned by the disposition to seek a common good involving people affected by the agreement. In these situations, the principles of equivalence in justification between two worlds are equated mutually, and a dispute's participants recognise their mutual contribution to a common good. As a result of the negotiations, an agreed text "comprises more than one order to worth in often fragile arrangements"[50]. The compromise blending different elements of liberal and sovereigntist claims will be reflected in the text's normative hybridity resonating with the respective orders of worth. Arrangements incorporating different normative stances create novel properties that might be reflected in normative configurations[51], complex norms[52], or norms' clusters[53].

Second, the compromise can also materialise by referring to ambiguous concepts and objects resonating with different orders of worth[54]. Actors accomplish a compromise by searching for precise consensual formulations and terms "acceptable to all - one that 'sounds right'"[55]. During negotiations, when the text is tested sentence after sentence, an amendment can expand the acceptable meaning of the text when the original wording is interpreted as too narrow. In this way, it aims at increasing the ambiguity of the text. Conversely, an amendment may specify the text's meaning by limiting its scope and narrowing it to specific normative order to decrease ambiguity. Ambiguous concepts, terms and wordings open to multiple interpretations are pervasive in global governance, given the inherent nature of language open to various interpretations, the subjectivity of understandings and the contingency of meanings depending on a multiplicity of contexts[56]. In a situation of normative multiplicity, amendments reinforcing the ambiguity of contentious elements will have better chances of being accepted than those restricting the text's meanings. Ambiguity offers a conceptual tool to capture multiple meanings and ambiguous concepts can be exploited as a governance practice emerging from indeterminacy and different interpretations[57]. So-called "constructive ambiguity" is frequently used in international negotiations to accomplish a compromise[58] and may lead to the emergence of normative ambiguity, which can favour flexible adjustments in norms implementation[59].

### 2.3. Data and methods

The empirical data analysed in this paper includes written documents produced while elaborating the text of the UNESCO AI Recommendations and the transcripts of video recordings of online negotiations during the restriction related to the COVID-19 pandemic. This textual material has been coded to distinguish relevant themes and concepts to determine the different normative positions and two types of compromises.

First, to map normative multiplicity, the methodology involves thematic coding of the transcripts of the opening statements presenting state priorities. I followed an abductive approach comprising the collection of pertinent empirical observations and, concurrently, applying concepts from Table 1 to substantiate initial expectations with empirical observations[60]. The first coding round focused on defining general state positions on AI technology as reflected in the draft text of UNESCO AI Recommendations. During the subsequent rounds of coding, I organised these observations following the analytical categories constituting an order of worth tested by states in their appreciation of the draft document (see Table 1). Second, I have coded the draft text of the Recommendation by searching for the markers of the structural normative hybridity reflected in key concepts relevant to two analysed orders of worth, in this case, focusing on the state's role and human rights. Third, I explored the debates on the proposed amendments to illustrate how negotiators searched for compromise by testing the draft from the perspective of different orders of worth. Available visual and textual data allowed for tracing online text drafting by employing the insights from the track-changes diplomacy approach[61].

## 3. The liberal-sovereigntist multiplicity

This section analyses the opening statements of 49 member states on the priorities towards the intergovernmental negotiations reflecting states' general views on AI technologies and the draft document. It illustrates that the elements of the liberal order of worth following the *principle of universalism* and the sovereignist order of worth constituted around the *principle of particularism* are frequently amalgamated in state positions connecting different elements.

The thematic analysis of state positions shows that many countries combined arguments from both orders of worth, reflecting normative multiplicity rather than clear-cut cleavages (Figure 1). Most countries adopted positions including references to both orders of worth, while only one state represented purely sovereigntist positions and eleven liberal ones. As illustrated by the trendline, there is no clear-cut cleavage representing liberal and sovereigntist values, but different distribution of priorities among them. While remarkably more countries referred to the elements of the liberal order of worth, sovereigntist elements still featured considerably in their positions. Overall, the analysis of statements shows that states referred twice more to the liberal than sovereigntist elements (227 arguments vs 113 arguments).

**Figure 1**. State positions on the negotiation of the UNESCO Recommendation of AI Ethics

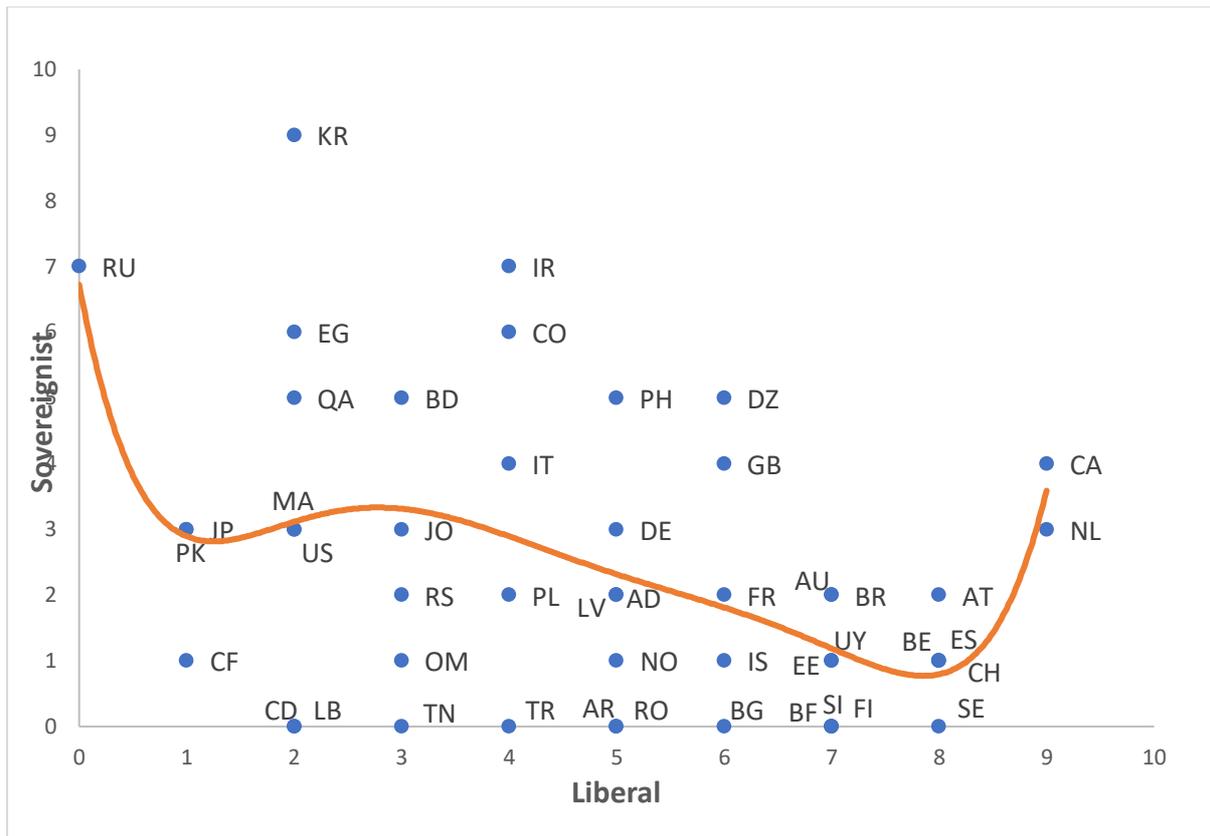

Source: Own elaboration

From the perspective of multiple normative positions, states, through the opening statements, offered the initial tests of the overall content of the draft text. States usually performed in opening statements truth tests reaffirming their overall support for the existing draft and frequently pointing out elements shared among different states. The emphasis on subtle differences in relation to shared elements signalled that during the negotiations, states would perform reality tests to modify the wording in the draft. The overall critique of the draft text reflecting the existential test was performed only by Iran and the Russian Federation, but they signalled their readiness to negotiate the text.

The first shared element of both orders is that the arrival of AI requires the *protection* of common goods and offers *opportunities* for common goods. Liberal positions focused on protecting individual freedoms, progress and development from the "threats to privacy to dignity and the personality of individuals" brought by AI (Burkina Faso) [62]. As argued by many states, AI should be "human centred and has human dignity as one of its core values" (Spain), while "human autonomy must be a central principle" (Germany). In turn, sovereignist stances also emphasise protection but focus on self-determination, cultural autonomy and state sovereignty against the challenges created by AI. States should have guaranteed their autonomy to implement the document since it "should offer interested states various possible solutions that there should not be any imposition of a unilateral kind of approach" (Russian Federation). At the same time, many states representing different normative positions emphasised that the benefits of AI systems are universal by contributing to the global agenda of Sustainable Development Goals, innovation, economic development, and social progress. Therefore, for the Netherlands, "it will be imperative to strike a positive balance between the risks and the benefits based on a clear sense of justice and fairness".

Secondly, in both cases, a common good is the *community*. A more liberal perspective focuses on humankind as a community. AI is a challenge "that will have a major impact on society at large, on human interactions, although the very experience of human nature, human autonomy and human agency" (Brazil). More sovereignist stances focus on communities constituted around states and cultures. For example, Qatar stressed that "religious wisdom is an integral part of people's moral world in different parts of the world" (Qatar). But even in the case of Canada, the representative of Quebec highlighted that the "promotion of the diversity of cultural expression are matters that we support very strongly".

Third, the focus on *ethical values* is the virtue emphasised across different orders of worth, even though they are framed around competing principles of universality and particularity. For more liberal views, the debate should focus on global commonalities rather than specificities and, therefore, "must develop a global consensus based on minimum ethical framework universally acceptable to all mankind, to prevent collisions between different ethical systems applied to AI by various stakeholders" (Bulgaria). More liberal states emphasised the need to guarantee the protection of universal legal commitments since "human rights are universal rights, they are indivisible and undeniable, they cannot be deemed optional when considering the provisions of our recommendation on AI" (Switzerland). However, for other states, recognising different ethical systems allows for accepting "the different starting points of each country and their respective priorities and circumstances" (Egypt). Therefore, more sovereignist countries did not neglect the development of a universal approach across state differences. They emphasised a different starting point for developing a shared perspective since there is a "need for strong AI ethics and governance framework rooted in a local context and aligned to international norms" (Qatar). Instead of looking at some common denominator, "each government should first acknowledge the importance of ethical issues caused by AI technology and build a broad social consensus on the ethics of AI through research and education" while "developing an action agenda that can be agreed by all countries should come after that" (South Korea).

Fourth, different states emphasised the problem of *discrimination, exclusion, and inequality* as the deficiencies to be addressed in the document. Liberal-oriented states emphasised that the situation of individuals requires guaranteeing the values of equality and inclusivity of persons. They underlined the priority of promoting gender equality to uproot this source of inequality and prevent discrimination, stigmas, and prejudices about women. As insisted by Austria, the instrument should "be a positive tool to transform unequal power relationships and promote human rights, in particular, gender equality", and to be "truly gender transformative", it should "address the underlying root causes for gender inequality". At the same time, however, more sovereignist-oriented states complained that cultural sensitivities are treated unequally since "some philosophical stances are overemphasised" as, for example, sexual orientation and gender identity "are controversial and may only stir up conflicting philosophical convictions between member states and their deeply rooted beliefs and cultures" (Algeria). Others emphasised the need to preserve the diversity of cultures, such as maintaining "the plurality of the natural languages we have worldwide that are a patrimony of humanity" (Italy).

Fifth, the *law* constituted the shared valued object linking the universal agenda to promote human rights and the sovereignist agenda of its context-sensitive implementation. In line with the universalist approach, universal human rights protection constituted the priority as the most *valued object*. The Recommendations should be "firmly rooted in the applicable international legal framework in international law, in particular international human rights law" (Austria). At the same time, sovereignist-oriented countries emphasised the value of national laws to implement this UNESCO framework document. For example, they stressed that digital technology companies with extraterritorial impacts should "fully respect national law and regulations and social and cultural values of different countries" (Iran).

The *state* constituted another valued object as the point of reference for different positions. The more liberal approaches indicated that even the universal agenda would not undermine states' responsibilities and particularities representing a variety of positions. For example, the Netherlands, attached to the global liberal approach, emphasised that while the recommendation provides "a normative framework and guidance", at the same time, "all states have to define their own appropriate rules but should do so inspired by and in tune with this global framework". In this context, sovereignist states emphasised the need to respect the state authority as regards the regulation of AI, given the responsibilities for order and security within their borders. They highlighted the non-binding character of the Recommendation since "no matter how strongly worded, would always remain a recommendation that is optional to implement" (Egypt). Similarly, Germany, bridging both perspectives, supported human rights standards as "the cornerstone of our work in general" but supported "a language that reflects the non-binding character of the recommendations".

Such a normative multiplicity with shared elements across different orders of worth created the context for negotiating the text of Recommendations. The following section shows that the document's draft already included many shared elements from both orders of words and created conditions for a compromise during the negotiations.

## 4. Towards a compromise: drafting the normative hybridity

The Recommendations' first draft text was prepared during consultations by an Ad Hoc Expert group in 2020[63]. This early text has been re-drafted by the UNESCO Secretariat after receiving comments from 50 member states in 2020 and early 2021. Therefore, in spring 2021 intergovernmental debates focused on the draft reflecting the opinions submitted before the formal process of intergovernmental negotiations[64].

**Table 2.** Words frequency in the UNESCON Recommendations draft

| Word | Final versions | | Draft version | | Similar words |
|---|---|---|---|---|---|
| | Count | Weighted Percentage | Count | Weighted Percentage | |
| AI | 347 | "3,81%" | 333 | "3,89%" | "#ai, ai" |
| systems | 194 | "2,13%" | 180 | "2,10%" | "system, systems" |
| states' | 140 | "1,54%" | 132 | "1,54%" | "state, states, states'" |
| humans | 140 | "1,54%" | 132 | "1,54%" | "human, humanities, humanity, humans, humans'" |
| member | 135 | "1,48%" | 125 | "1,46%" | "member, members" |
| ethical | 108 | "1,19%" | 108 | "1,26%" | "ethical, ethically, ethics" |
| rights | 92 | "1,01%" | 89 | "1,04%" | "right, rights" |
| technologies | 90 | "0,99%" | 95 | "1,11%" | "technological, technologically, technologies, technology" |
| ensure | 90 | "0,99%" | 78 | "0,91%" | "ensure, ensured, ensuring" |
| development | 88 | "0,97%" | 83 | "0,97%" | "develop, developed, developers, developing, development, developments" |
| data | 87 | "0,96%" | 85 | "0,99%" | data |

| international | 74 | "0,81%" | 68 | "0,79%" | "international, internationally" |
| using | 66 | "0,72%" | 53 | "0,62%" | "use, used, uses, using" |
| promote | 64 | "0,70%" | 59 | "0,69%" | "promote, promoted, promoting, promotion" |
| impact | 63 | "0,69%" | 57 | "0,67%" | "impact, impacts" |
| policy | 63 | "0,69%" | 53 | "0,62%" | "policies, policy" |
| life | 60 | "0,66%" | 57 | "0,67%" | life |
| research | 58 | "0,64%" | 53 | "0,62%" | "research, researchers" |
| cycle | 55 | "0,60%" | 53 | "0,62%" | cycle |
| protection | 53 | "0,58%" | 49 | "0,57%" | "protect, protected, protecting, protection" |

Source: Own elaboration

From the structural perspective, the content analysis of the draft document shows how key concepts relevant to liberal and sovereigntist orders of worth feature equally in the text. While "AI systems" are the point of reference in the draft document, "states" and "human rights" appear as the most frequently and equally employed terms. The frequency shows that states' driving role in addressing concerns related to human rights is merged in the text. As such, this hybridisation was a deliberate effort of the UNESCO Secretariat since the draft text "recognises the existence of different ethical systems along cultural diversity" but stresses that "at the same time, the goal of UNESCO (…) is to find the ethical common ground that will allow the international community to advance in achieving the responsible use of AI technologies within the UN human rights framework"[65].

The draft document offered scope for flexible interpretation and implementation, merging "a universal framework of values, principles and actions" to guide "States in the formulation of their legislation, policies or other instruments regarding AI, without prejudices to existing international law"[66]. Furthermore, it also blended liberal and sovereigntist priorities since the document aimed at guiding the actions of "individuals, groups, communities, institutions and private sector companies to ensure the embedding of ethics in all stages of the AI system life cycle" as well as "to protect, promote and respect human rights and fundamental freedoms, human dignity and equality, including gender equality (…) and to respect cultural diversity in all stages of the AI systems life cycle"[67].

In line with liberal priorities, the document mentioned 63 times the need to respect human rights and 25 times the respect for international law. It already included a preliminary list of non-discriminatory criteria. The draft defined the values of respect, protection and promotion of human rights, fundamental freedoms, and human dignity; environment and ecosystem flourishing; ensuring diversity and inclusiveness; living in peaceful, just, and interconnected societies. These values are further unpacked in principles of proportionality and do no harm; safety and security; fairness and non-discrimination; sustainability; right to privacy and data protection; human oversight and determination; transparency and explainability; responsibility and accountability; awareness and literacy; and multi-stakeholder and adaptive governance and collaboration.

At the same time, however, this document emphasises that states are the main subjects responsible for implementing the document both as "AI actors and as authorities responsible for developing legal and regulatory framework throughout the entire AI system life cycle, and for promoting business responsibility"[68]. To operationalise the values and principles of the Recommendations "the main action is for Member States (…) to ensure that other stakeholders, such as private sector companies, academic and research institutions, and civil society adhere to them"[69]. Of 80 paragraphs covering the 11 policy areas, only two do not explicitly focus on the responsibility of Member States concerning the regulation of AI in policy domains[70]. States emerge as the central loci of authority as they should "ensure", "encourage", "foster", "carry out", or "develop" specific actions and regulations. As a Recommendation, this document is legally non-binding on member states, but it should be applied "to give effect within their jurisdictions to the principles and norms of the Recommendation in conformity with international law, including international human rights law"[71].

## 5. Text negotiations as normative testing

This section interprets the final stage of negotiations as the process of testing the meanings of the draft text from multiple normative perspectives. The proposed amendments tended to reaffirm, enhance, or contest the match between specific text wordings and the orders of worth defining states' positions. They converted the negotiation into a collective process of drafting a text by debating the wording to accommodate different perspectives.

The negotiations of the draft text in April and June 2021, supported by the UNESCO Secretariat, comprised the consecutive discussion of each paragraph of the document[72]. The paragraphs without amendments were approved without any debate, but in most of the cases, states proposed amendments. To streamline the debates, member states were asked to submit their amendments to the UNESCO Secretariat, which would assess them before admitting them to plenary[73]. If admitted, the Secretariat would read the paragraph with proposed amendments visible on the screen, and the Chair would ask if there were any objections before confirming the approval. If there was any objection to the amendment, the Chair opened the debate on the proposed amendments. The debate on each amendment led to its acceptance, further modifications, or refusal. Overall, states proposed around 500 amendments to the draft text. The analysis of the most contentious issues during the negotiations (see Annex 1) shows that the debates on related amendments mirror the three types of tests conceptualised by Boltanski. The outcome of testing each amendment depended on the scope of their perceived ambiguity. The amendments reflecting situated normative ambiguity linking different orders of worth had better chances of being accepted than those perceived as supporting one perspective.

Many amendments proposed during the negotiations went through *truth tests* since it was recognised that their wording only reinforced their embedding within different orders of worth.

In these situations, the acceptance was straightforward when it was possible to justify them by different principles. For example, the inclusion of the two words in the first paragraph (comprehensive, multicultural), exemplifying respect for both different cultures and universal scope, was accepted since it was recognised by different states as fitting with the principles of universality and particularity (Figure 2). As explained by Germany, "it is quite self-understanding for us, even though it does not seem to have substance there", but the concept of "multicultural" already drew from the language used by UNESCO[74].

**Figure 2.** Example of amendment after truth test

> "(…) It approaches AI ethics as a systematic normative reflection, based on a holistic, [Iran -accepted] *comprehensive, multicultural,* and evolving framework of interdependent values, principles and actions that can guide societies in dealing responsibly with the known and unknown impact of AI technologies on human beings, [Iran – refused after debate] *respect to cultural diversity, different values systems of* societies, and the environment and ecosystems, and offers them a basis to accept or reject AI technologies"

Source: Own elaboration

Other analysed truth tests involved the substitution of "constitutions" with "constitutional practice" and including "on a voluntary basis" in Preamble 26 to reflect different states' practices regarding the implementation of the Recommendations. Similarly, for the sake of flexibility and given the difficulty of being comprehensive in the Preamble, states refrained from listing entities, organisations, human rights, legal principles, and political documents.

Many sections of the draft went through *reality tests* criticising the existing draft wording and proposing an amendment to enhance its resonance with relevant orders of worth. The debates show that many substantive amendments were proposed to accommodate different normative positions. For example, the amendments to the paragraph regarding the responsibility and accountability principle concerning human rights and fundamental freedoms involved a simultaneous debate on 13 amendments proposed by eight countries (Figure 3).

**Figure 3.** Example of modified amendments after reality tests

42. [UK – substitution not accepted] *Member States have an obligation to* ~~AI actors~~ [St Lucia – compromise accepted] AI actors *and Member States* [Australia – substitution accepted] ~~must~~ *should* respect, protect, and promote [Singapore – addition refused without debated] *applicable* human rights and fundamental freedoms, and should [UK – accepted without debate] *also* promote the protection of the environment and ecosystems, assuming [Germany – addition accepted after debate] *their respective* ethical and [South Korea – deletion refused without debated] ~~legal~~ responsibilities [Hungary – addition accepted]*,* [Bangladesh – substitution not accepted] *following* ~~in accordance with~~ [UK – substitution not accepted] *in line with* ~~in accordance with~~ [Singapore – substitution not accepted] ~~existent~~ *applicabl*e [Canada – deletion accepted] ~~existent~~ national and international law, [UK – initially to delate then to keep, accepted to keep after debate] *in particular Member States' human rights obligations*, and ethical guidance throughout the life cycle of AI systems [UK – addition accepted without debate], *including with including with respect to AI actors within its effective territory and control.* The ethical responsibility and liability for the decisions and actions based in any way on an AI system should always ultimately be attributable to AI actors [Canada – accepted without debate] *corresponding to their role in the life cycle of the AI system*.

Source: Own elaboration

During this discussion, 12 states engaged in the collective drafting of the text to balance its universal application and the particular role of the states in its implementation. The initial UK proposal to replace "AI actors" by "Member States" was justified to "clarify" and "explicitly" introduce this distinction between states' "obligations" and actors' "responsibilities". However, it was contested, since Brazil, Iran, Egypt, and Cuba claimed that the obligations regarding human rights should be universal and not limited to member states. Therefore, to find a compromise, Santa Lucia, supported by Germany, Poland, and Mexico, proposed to include both "AI actors" and "Member States" in the opening of the sentence, being the first part flexible and open to broader interpretation, but also emphasising the unique role of Member States.

Given this compromise, the United Kingdom, supported by Germany, Canada, and Brazil, after suggesting deleting "in particular Member States' human rights obligations", proposed to keep it, claiming that it both had broad transversal support and it clarified the norm. Nevertheless, this debate led Australia to suggest replacing the word "must" by "should" as it would limit the compulsory character of the prescription concerning the generic "AI actors". The United Kingdom and Santa Lucia supported this change appealing that it would clarify the text from a legal-technical perspective and be consistent with the language used in other parts of the Recommendations. At the same time, however, to avoid possible confusion and limit ambiguity, Germany argued about the inclusion of "their respective" to differentiate the obligation of member states and the responsibilities of other actors and, in so doing, to clarify the text[75]. As a result, the process of collective drafting with inputs from different normative perspectives rebalanced the scope of ambiguity but still allowed for accommodating different interpretations guiding subsequent actions regarding the role of states and other actors in relation to human rights.

Similar time-consuming debates focused on searching for wording that would meet different normative positions, consequently leading to more ambiguous understandings. For example, it was decided to exclude from the definition of AI systems "technological" or "socio-technological" qualifications; leave flexible the ground for non-discrimination by including a catchphrase "any other grounds" instead of offering a closed catalogue of discriminatory grounds; clarify the voluntary character of assessment, monitoring and evaluations by UNESCO; underline holistic nature of development rather than its conception as a human right; and, consider the ethical reflection as subsidiary to norms "development" rather than norms "harmonisation".

*Existential tests* aimed at subverting the negotiated text by introducing amendments explicitly promoting one specific normative perspective. For example, the debate on the Iranian amendment in paragraph 41 illustrates that states refused it given that it would significantly narrow its practical application (Figure 4). Iran justified this proposal very briefly: "This is just an emphasis on the role that international corporations can play concerning their responsibilities"[76].

**Figure 4.** Example of refused amendment after the existential test.

> 41. Transparency and explainability relate closely to adequate responsibility and accountability measures, as well as to the trustworthiness of AI systems, [Iran – refused after debate], *especially with regard to platforms and transnational corporations*.

Source: Own elaboration

The United Kingdom, Canada, France, and Argentina argued against this inclusion, contesting that this specification would limit its universal applicability and flexible adaptation to different circumstances. It was understood that Iranian amendments aimed at circumventing the activities of multinational companies and reinforcing state sovereignty. Kuwait and Bangladesh also preferred to keep the original text, stating that it is "sufficient to the purpose" and "good enough" without elaborating more on their position. As a result, the Chair considered that the original text should be approved without any amendments[77].

Iran and Russia usually proposed many amendments questioning the draft text. They were refused, given that it was understood that specific amendments would limit other states' scope of actions and the scope of interpretation of the Recommendation. Some amendments aimed to include the reference to national legislation on open data; to add "extra-territorial impact" regarding private sector companies operating at the international level; to exclude the terminology of "whistle-blower protection", "gender diversity", "precedence of human rights" and the reference to "local customs and religious traditions". During these debates, states argued that these amendments would limit the flexible understandings that accommodate different positions.

**6. Conclusion**

The adoption of the UNESCO Recommendations on the Ethics of AI by 193 states – the first global instrument regulating AI – concluded after prolonged debates on more than 500 amendments. Despite its non-binding character, after one year of preparatory work, UNESCO member states spent five weeks of intense negotiation to reach a compromise accommodating multiple preferences merging the liberal and sovereigntist positions[78]. To understand how member states reached a compromise, this paper offers an elaborated conceptualisation of the practice of multilateral negotiating drawing from the work of Luc Boltanski.

Analysing the orders of worth constituted around the liberal and sovereigntist references shows that many states adopted an amalgam of positions. The analysis of state positions reflected that they shared the adherence to such common elements as protecting/benefiting from AI, community, values, non-discrimination, law, and the state. Even though these positions were qualified in different ways drawing from liberal and sovereigntist positions, they constituted the points of reference for the states negotiating positions. Therefore, given the normative multiplicity, no stable coalitions were confronted during the negotiations, but ad hoc pragmatic alignments concerning specific issues. The compromise could be reached given the two mechanisms which bridged normative differences between participating states.

First, the structural normative hybridity included in the draft text reflected a balance between key claims inspired by liberal and sovereigntist positions. It shows the relevance of the UNESCO Secretariat as an intermediary actor able to synthesise different states' views and skilfully blend in the draft text the references to the universal protection of individual human rights and the particular role of states in implementing AI policies. Second, during the intergovernmental negotiations, states conducted, through 500 amendments, multiple tests of the draft text to reaffirm, reinterpret, or criticise the consonance of the most contentious issues with multiple normative perspectives. This compromise mechanism of searching for the situated normative ambiguity avoided a clash of incommensurable orders of worth blocking the negotiations based on the unanimity rule and allowed to refuse terminology which would constrain flexible interpretations.

This paper also offers future research avenues. First, it can be expected that the apparently universal framework of norms on AI will be localised[79], leading to the fragmentation of the AI regulations since states will prioritise those elements more in line with their approaches. Therefore, it can be studied how the mechanisms of hybridisation and ambiguity might lead to selective implementation and fragmentation within the landscape of normative multiplicity. As a result, it remains to be seen whether this global document will indeed lead to the emergence of a universal framework of AI ethics at the level of state regulations. Second, the conceptual framework to study macro-normative structures in situated micro-interactions allows for a nuanced understanding of liberal-sovereigntist antagonism and the mechanisms to achieve compromises within multilateral frameworks. It can be adapted to different international settings to understand how states overcome disagreement. Multilateral negotiations bringing novel global agreements such as the UN High Seas Treaty adopted from March 2023 and the Global Biodiversity Framework adopted in December 2022 exemplify that states might reach global agreements despite very entrenched preferences and a complex process of negotiations.

**Notes**

[1] O'Neill, *Weapons of Math Destruction.*

[2] Amoore, *Machine learning*; Dragu and Lupu, "Digital Authoritarianism"; Feldstein, *The Rise of Digital*; Noble, *Algorithms of Oppression*; Aradau and Blanke, *Algorithmic Reason*.

[3] Schmitt, "Mapping global AI".

[4] UNESCO, "Preliminary study", 35-36.

[5] UNESCO. *Final Report.*

[6] Even though Iran and Russia registered several reservations about various important aspects of the document, and the United Kingdom, Singapore, Poland, and Morocco requested to reflect some specific concerns, they did not block the final consensus. UNESCO. *Final Report Intergovernmental meeting.*

[7] Zürn, *A Theory of Global Governance*.

[8] Weiss and Wallace, "Domestic Politics, China's Rise"; Paris, "The Right to Dominate".

[9] Huang, "The Pandemic"; Adler-Nissen and Zarakol, "Struggles for Recognition".

[10] Zürn and De Wilde, "Debating globalization"; De Wilde et al. *The Struggle Over Borders*.

[11] Flonk, Jachtenfuchs, and Obendiek, "Authority conflicts".

[12] Müller, "Arguing, Bargaining and All That", 401.

[13] Shelton, "Introduction", 4.

[14] Shelton, "Soft Law".

[15] Risse, "Let's Argue!"; Deitelhoff, "The Discursive Process of Legalization", Ulbert and Risse, "Deliberately Changing the Discourse"; Deitelhoff and Müller, "Theoretical Paradise".

[16] Fehl, "Bombs, Trials, and Rights"; Moore, "Negotiating Adaptation"; Pratt, "From norms to normative configurations"; Lesch, "Multiplicity, hybridity", Winston, "Norm structure"; Byers, "Still agreeing to disagree"; Widmaier and Luke Glanville, "The benefits of norm ambiguity"; Linsenmaier, Schmidt and Spandler, "On the meaning(s) of norms".

[17] Obendiek, "What Are We?".

[18] Reckwitz, "Toward a Theory"; Pouliot, "The Logic of Practicality".

[19] Pouliot, *International Pecking Orders*, 49.

[20] Bueger and Gadinger, *International Practice Theory*, 26-30.

[21] Boltanski and Chiapello, *The New Spirit*; Boltanski and Thévenot, *On Justification*. This approach has been employed to study controversies on the US war on terror, massive surveillance disclosed by data governance, Edward Snowden, global health policies, and UN Security Council debates about interventions and corruption. Obendiek, "What Are We?"; Gadinger, "On justification". Ochoa, Gadinger and Yildiz. "Surveillance under dispute"; Hanrieder, "Orders of worth"; Niemann, *The Justification of Responsibility*; Lesch, "Multiplicity, hybridity".

[22] Bueger and Gadinger, *International Practice Theory*, 89-90; Franke and Weber, "At the Papini Hotel", 657-677.

[23] Joas, *The Creativity of Action*, 154-161.

[24] Hanrieder, "Orders of worth", 398.

[25] Lesch, "Multiplicity, hybridity", 617.

[26] Boltanski, *On Critique*, 27.

[27] Boltanski and Thévenot, *On Justification*, 66.

[28] Hanrieder, "Orders of worth".

[29] Hanrieder, "Orders of worth", 398.

[30] Bueger and Gadinger, *International Practice Theory*, 89.

[31] Boltanski and Thévenot, *On Justification*.

[32] Boltanski and Chiapello, *The New Spirit*.

[33] Niemann, *The Justification of Responsibility*, 236; Hanrieder, "Orders of worth".

[34] Lake, Martin and Risse, "Challenges to the Liberal Order".

[35] Flonk, Jachtenfuchs, and Obendiek, "Authority conflicts in internet governance".

[36] Zürn and De Wilde, "Debating globalization"; De Wilde et al. *The Struggle Over Borders*.

[37] Bueger and Gadinger, *International Practice Theory*, 88.

[38] Boltanski, *On Critique*: 25, emphasis in original)

[39] Ibid, 133.

[40] Boltanski, *On Critique*.

[41] Boltanski, "A Journey Through", 44.

[42] Pouliot, "Multilateralism as an End", 9.

[43] Hanrieder, "Orders of worth", 414.

[44] Niemann, T*he Justification of Responsibility*.

[45] Boltanski and Thévenot, *On Justification*, 215.

[46] Boltanski and Thévenot, *On Justification*, 278.

[47] Hanrieder, "Orders of worth", 412.

[48] Adler-Nissen and Drieschova, "Track-Change Diplomacy", 537.

[49] Ibid, 277.

[50] Lesch, "Multiplicity, hybridity", 614.

[51] Pratt, "From norms to normative configurations".

[52] Fehl, "Bombs, Trials, and Rights".

[53] Winston, "Norm structure".

[54] Boltanski and Thévenot, *On Justification*, 279-280.

[55] Boltanski and Thévenot, *On Justification*, 281.

[56] Best, "Ambiguity, Uncertainty", 362.

[57] Best, "Ambiguity and Uncertainty in International Organizations"; Best, "Bureaucratic ambiguity."

[58] Byers, "Still agreeing to disagree".

[59] Widmaier and Luke Glanville, "The benefits of norm ambiguity"; Linsenmaier, Schmidt and Spandler, "On the meaning(s) of norms".

[60] Kratochwil and Friedrichs, "On Acting and Knowing".

[61] Adler-Nissen and Drieschova, "Track-Change Diplomacy", 539-540.

[62] UNESCO, *Video transcripts*, 26 April 2021. All subsequent state quotes are from this source.

[63] UNESCO. *Preliminary Report*.

[64] UNESCO. *Draft text*, 31 March 2021. UNESCO. *Final Report*, 31 March 2021.

[65] UNESCO. *Final Report*, 31 March 2021, 6.

[66] UNESCO. *Draft text*, 31 March 2021, 6.

[67] Ibid.

[68] Ibid, 6.

[69] Ibid, 12.

[70] The policy areas are ethical impact assessment, ethical governance and stewardship, data policy, development and international cooperation, environment and ecosystems, gender, culture, education and research, communication and information, economy and labour, health, and social well-being.

[71] UNESCO. *Draft text*, 31 March 2021, 4.

[72] UNESCO. *Draft text, Paragraphs 1-25*; UNESCO. *Draft text, Paragraphs 26-131*.

[73] The criteria to admit the amendments to debates were: 1) high importance for the countries; 2) focus on the substance of the Recommendation, 3) preservation of the effectiveness of the non-binding instrument, 4) avoidance of redundancy by raising the issues already debated or that might be addressed elsewhere in the text, and issues regarding specific language, style, and other details not related to the substance of the instrument. UNESCO. *Final Report by the Rapporteur,* 2; UNESCO. *Video transcripts.*

[74] UNESCO, *Video transcripts*, 26 April 2021.

[75] Ibid.

[76] Ibid.

[77] Ibid.

[78] Zürn and De Wilde, "Debating globalization".

[79] Acharya, "How Ideas Spread".

**Annex 1. Text amendments and ambiguity**

|   | Proposed amendments | Amendment justification | Outcome of debate | Outcome justification |
|---|---|---|---|---|
| 1 | Avoid long listing entities and documents in the text | Difficult to be comprehensive | Accepted | More flexibility |
| 2 | PP 26 – add "on a voluntary basis" | Emphasize non-binding character | Accepted | More flexibility |
| 3 | PP 26 – substitute "constitutions" with "constitutions practice" | Reflect that some countries have no written constitutions | Accepted | More flexibility |
| 4 | PP 27 – the role of stakeholders | Soften the compulsory character | Accepted | More flexibility |
| 5 | Substitute AI as "technological" with "socio-technological" systems (Paragraph 2) | Broadening to include the social impact of AI | Compromise - "AI systems" | Expands ambiguity by including the broadest range of systems |
| 6 | Paragraphs 13, 19, 28, 74. Adapt the list of non-discrimination grounds | Restrict the scope by avoiding "controversial" categories | Compromise - redefined list and includes "any other grounds" | Expands ambiguity by offering flexibility in exchange for modifying the draft text criteria |
| 7 | Paragraphs 49, 58, 131, 132. Redefine the Ethical Impact Assessment and readiness assessment methodology | Limit the role of UNESCO and reduce the emphasis of assessments | Compromise – addition of "interested states" | Increased ambiguity so no interference and allows flexibility of UNESCO mandate to support states |
| 8 | Multiple paragraphs. Include "the Charter of the United Nations" when international law is mentioned | Specify international law | Compromise included in Para 9 and Preamble 4 | Retains ambiguity and flexible scope of international law. |
| 9 | Preamble. Right to development | Specify that development has human rights status | Compromise – Mention in preamble without status of rights the inclusion of the United Nations resolution | Allows the holistic meanings of development, including sustainable aspect, and human rights |
| 10 | PP 13 Relation between laws and ethics | Clarify superiority of laws against ethical guidance | Compromise – flexible possibility of ethical contribution to rights-based policy measures and legal norms | Allows for flexible inclusion of ethical reflection in the development of laws |
| 11 | PP14 Harmonizing AI-related norms | No mandate for normative harmonization | Compromise – instead of "harmonization" "development" | Expands flexibility– beneficial to have a variety of legal norms |
| 12 | Preamble 22; Paragraph 51. Access to information is not recognized as a right | Exclude some categories from the terminology of rights | Accepted - access to information is not a right | Expands flexibility as regards the nature of the access to information |
| 13 | Paragraph 14. Harm to human beings and communities' definition | Specify the distinction | Refused | Retains ambiguity since the approved original form is a good balance |
| 14 | Paragraph 43. Remove "whistle-blower protection" | Delete concept since no national legislation exists in some cases | Refused | Original allows flexibility in other countries |
| 15 | Paragraph 106. Include "minatory languages" in AI ethics education | Specify the scope of protection | Refused | Original is ambiguous enough since "local languages, including indigenous languages" implicitly includes other notions |
| 16 | Paragraph 61. Add "extra-territorial impacts" after "the private sector" | Specify the responsibility and expanding state prerogatives | Refused | Limits ambiguity by excluding other actors |
| 17 | Paragraph 75. Add national legislation in relation to open data | Specify the scope of the framework | Refused | Limits ambiguity and it is implicitly understood that legislation will be respected |
| 18 | Paragraph 92. Substitute the "gender diversity" by "gender equality" or "boys and men" | Specify the nature of gender | Refused | Limits ambiguity of the flexible meaning of gender policies |
| 19 | Paragraph 65. Delete "precedence of human rights" and reference to "local customs and religious traditions" | Reinforce the central role of the state | Refused | Original includes ambiguity balancing between social diversity and human rights |